\documentclass[twocolumn,trackchanges]{aastex631}

\hypersetup{linkcolor=red,citecolor=blue,filecolor=cyan,urlcolor=magenta}

\usepackage{graphicx}

\usepackage{newtxtext,newtxmath}
\usepackage{amsmath}
\usepackage{amssymb}
\graphicspath{{figs/}}
\usepackage{algorithm} 
\usepackage{algpseudocode}
\usepackage[toc,page]{appendix}
\usepackage{comment}
\usepackage[normalem]{ulem}
\usepackage{soul}

\graphicspath{{./}{figs/}}
\begin{document}

\title{An Unsupervised Learning Method for Radio Interferometry Deconvolution}

\correspondingauthor{Lei Yu}
\email{yulei@nao.cas.cn}

\correspondingauthor{Bin Liu}
\email{bliu@nao.cas.cn}

\correspondingauthor{Bo Peng}
\email{pb@nao.cas.cn}

\author{Lei Yu}
\affiliation{CAS Key Laboratory of FAST, National Astronomical Observatories, Chinese Academy of Sciences, Beijing 100101, China}
\affiliation{University of Chinese Academy of Sciences, Beijing, 100049, China
}

\author{Bin Liu}
\affiliation{CAS Key Laboratory of FAST, National Astronomical Observatories, Chinese Academy of Sciences, Beijing 100101, China}

\author{Cheng-Jin Jin}
\affiliation{CAS Key Laboratory of FAST, National Astronomical Observatories, Chinese Academy of Sciences, Beijing 100101, China}

\author{Ru-Rong Chen}
\affiliation{CAS Key Laboratory of FAST, National Astronomical Observatories, Chinese Academy of Sciences, Beijing 100101, China}

\author{Hong-Wei Xi}
\affiliation{CAS Key Laboratory of FAST, National Astronomical Observatories, Chinese Academy of Sciences, Beijing 100101, China}

\author{Bo Peng}
\affiliation{CAS Key Laboratory of FAST, National Astronomical Observatories, Chinese Academy of Sciences, Beijing 100101, China}
\affiliation{Department of Astronomy and Institute of Interdisciplinary Studies, Hunan Normal University, Changsha, Hunan 410081, China}

\collaboration{20}{(AAS Journals Data Editors)}






\begin{abstract}
Given the incomplete sampling of spatial frequencies by radio interferometers, achieving precise restoration of astrophysical information remains challenging. To address this ill-posed problem, compressive sensing(CS) provides a robust framework for stable and unique recovery of sky brightness distributions in noisy environments, contingent upon satisfying specific conditions. We explore the applicability of CS theory and find that for radio interferometric telescopes, the conditions can be simplified to sparse representation. {{Building on this insight, we develop a deep dictionary (realized through a convolutional neural network), which is designed to be multi-resolution and overcomplete, to achieve sparse representation and integrate it within the CS framework. The resulting method is a novel, fully interpretable unsupervised learning approach that combines}} the mathematical rigor of CS with the expressive power of deep neural networks, effectively bridging the gap between deep learning and classical dictionary methods.  {{During the deconvolution process, the model image and the deep dictionary are updated alternatively.}} This approach enables efficient and accurate recovery of extended sources with complex morphologies from noisy measurements. Comparative analyses with state-of-the-art algorithms demonstrate the outstanding performance of our method, i.e., achieving a dynamic range (DR) nearly 45 to 100 times higher than that of multiscale CLEAN (MS-CLEAN).

\end{abstract}



\section{Introduction} \label{sec:intro}

With the construction and progressive deployment of advanced instruments such as the Square Kilometre Array (SKA), astronomers will benefit from improvements in sensitivity and angular resolution, facilitating the discovery of new physical phenomena. These advancements drive the ongoing development of novel imaging techniques. In the realm of radio synthesis imaging, deconvolution stands as an essential procedure, while generating images from interferometric data. Due to the incomplete sampling of spatial frequencies by the telescope array, image recovery is inherently an ill-posed problem. Deconvolution algorithms serve to restore the true sky brightness distribution from the calibrated visibility data. Among these algorithms, CLEAN and its variants (such as \cite{Hgbom1974APERTURESW,cornwell_multiscale_2008,schwab_relaxing_1984}) reign as the most popular and widely utilized methods in the radio community.

Compressed sensing theory, also known as compressive sensing (\cite{candes_stable_2006,donoho_compressed_2006}), refers to the problem of reconstructing a signal from a restricted set of measurements.  The essence of this methodology lies in acquiring fewer measurements than the Nyquist-Shannon sampling theory dictates, yet still being able to reconstruct the signal effectively. 

In the context of ill-posed radio interferometric imaging, addressing these challenges requires regularization or additional constraints to obtain meaningful solutions \citep{colton_ill-posed_2019}. CS  fundamentally provides a robust theoretical framework for the stable and unique recovery of sky brightness distributions, contingent on meeting key conditions such as sparsity, incoherence, and restricted isometry property (RIP) \citep{candes_introduction_2008}.

Numerous studies have undertaken efforts to apply CS techniques to radio interferometric imaging. For instance, \cite{li_application_2011} introduces two CS-based deconvolution methods well-suited for radio astronomy, which are Partial Fourier (PF) and Isotropic Undecimated Wavelet Transform (IUWT) based deconvolution.  Their experimentation is based on the Australian Square Kilometer Array Pathfinder (ASKAP) radio telescope. The PF method is effective for point sources, leveraging an $\ell_1$-norm based partial Fourier reconstruction. In contrast, the IUWT-based CS method is suitable for extended sources. This approach employs the IUWT as a sparsifying transform to represent the sky brightness distribution. 

Make use of both simulated and actual LOFAR data, \cite{garsden_lofar_2015} implemented a sparse reconstruction algorithm. Their investigation delineates three key aspects of sparse reconstruction performance. Firstly, leveraging wavelets and curvelets transformations, the method matches the performance of the CLEAN-based approach in recovering the flux of point sources and outperforms it for extended objects. Moreover, the method yields images with an effective angular resolution 2-3 times superior to the Cotton-Schwab CLEAN (CoSch-CLEAN, \cite{schwab_relaxing_1984}) method.

\cite{dabbech_moresane:_2015} presented a deconvolution algorithm named MORESANE, applied on the Karoo Array Telescope (MeerKAT). This algorithm is designed to restore faint, diffuse astronomical sources that may be obscured by the sidelobes of brighter sources within the same field. MORESANE employs a greedy algorithm that integrates both synthesis-based and analysis-based sparse recovery methodologies. The algorithm operates iteratively and does not rely on global optimization procedures. However, its performance is not necessarily superior to that of the sparsity-averaging reweighted analysis (SARA) \citep{carrillo_sparsity_2012} algorithm when applied to extended sources with complex morphologies. SARA assumes that the signal can be represented in a sparse domain. 

\cite{akiyama_superresolution_2017} conducted simulations of M87 observations using the Event Horizon Telescope (EHT) and utilized $\ell_1$ + total variation (TV) regularization. This approach significantly enhanced image fidelity and achieved an optimal resolution of approximately 25\%–30\% of the diffraction limit. 

\cite{cai_uncertainty_2018} developed a Bayesian inference-based deconvolution method that also incorporates sparsity priors as constraints, enabling both unique signal reconstruction and the quantification of uncertainties. \cite{muller_dog_hit_2022} introduced a multiscalar wavelet imaging algorithm designed for sparse representation, which is particularly well-suited for handling high-level sidelobes, as encountered in the EHT applications.



In recent years, deep learning methods have demonstrated exceptional performance across numerous fields, including radio interferometry. Many previous studies have applied supervised deep neural networks to perform deconvolution (e.g., 
\cite{nammour_shapenet_2022,schmidt_deep_2022,chiche_deep_2023,geyer_deep-learning-based_2023,liaudat_scalable_2024}), capitalizing on the powerful representational capabilities of these networks to enhance imaging quality and accuracy. For an overview of these works, refer to Appendix \ref{sec:dlm}. 

However, the deep learning methods listed are supervised models, which necessitate vast amounts of labeled data to perform effectively, incurring significant costs. The quality and quantity of training data, both obtained and simulated, are critical factors, making it challenging to encompass all scenarios. Moreover, these deep neural networks (DNNs) often act as "black boxes", making it difficult to interpret their decision-making processes \citep{lecun_deep_2015}.

In the context of $\ell_1$-norm sparse recovery problems, non-learning algorithms, such as the iterative soft-thresholding algorithm (ISTA), can be solved through convex optimization and do not rely on training data. Conversely, deep learning methods possess powerful expressive capabilities that can capture critical features of the data. Several works have attempted to integrate the advantages of traditional algorithms and deep learning techniques. For instance, \cite{zhang_ista-net:_2018} introduced ISTA-Net, a deep learning architecture inspired by the Iterative Shrinkage-Thresholding Algorithm (ISTA) for image reconstruction. \cite{han_tensor_2020} and \cite{xiang_fista-net:_2021} extended this concept by unfolding the Fast Iterative Shrinkage-Thresholding Algorithm (FISTA) into a deep neural network for video and image recovery.

On the other hand, these approaches typically employ CNNs as non-linear projections to represent the signal. This makes it challenging to interpret how the dictionary represents signals during both decomposition and reconstruction. Additionally, these methods primarily focus on supervised learning, which is costly and difficult to implement for individual users in the field of radio interferometry.

In this paper, we introduce a new deconvolution method for recovering the brightness distribution in radio interferometric imaging. This method incorporates \(\ell_1\)-norm and TV regularization as constraints. While our approach shares certain similarities with the work of \cite{akiyama_superresolution_2017} in terms of the selected constraints, it differs in several significant aspects. First, their method assumes the reconstructed image itself is sparse, whereas we leverage a deep dictionary to achieve a sparse representation, enabling broader applicability through adaptive sparse modeling. Second,  they use training and validation sets to determine regularization parameters, while our method employs unsupervised learning, which optimizes these parameters through an iterative process.  In Sect.\,\ref{sec:revisitc}, we revisit how radio interferometric telescopes satisfy the recovery conditions required by CS and propose a deep dictionary framework to enhance sparse representation. The Split Bregman method and its integration with our deep dictionary are detailed in Sect.\,\ref{sec:proposedm}. Comparisons between our deconvolution method and other methods are presented in Sect.\,\ref{sec:experimentalr}. Conclusions and discussions are given in Sect.\,\ref{sec:conclusion}.


\section{Revisit compressed sensing theory in radio interferometry\label{sec:revisitc}}

This section opens with an exploration of the applicability of CS theory to radio interferometric deconvolution. For most radio interferometers, the conditions under which CS theory is applicable can be simplified to a sparse representation. To enhance the capabilities of classical dictionaries (e.g., wavelet bases), we identified four key principles essential for achieving effective sparse representations. Building on these principles, we propose a deep dictionary for sparse representation.

\subsection{The Applicability of Compressive Sensing to Radio Interferometric Imaging}

Since radio interferometry samples partial spatial frequencies, the consequential deconvolution problem is inherently ill-posed. According to Hadamard, a well-posed problem must satisfy three criteria: the existence of a solution, the uniqueness of that solution, and its continuous dependence on the input data. When any of these conditions are not met, the problem is considered ill-posed, leading to an infinite set of possible solutions without the application of additional constraints\citep{colton_ill-posed_2019}. To address this challenge, CS can ensure stable and unique recovery of the original signal in noisy environments, provided that conditions such as sparsity, incoherence, and the restricted isometry property (RIP) are satisfied. CS establishes a relationship between the number of measurements $ m $, the signal dimension $ n $, and the $ k $ most significant coefficients, which can inform observational strategies. For a brief introduction to CS theory and the key conditions, see Appendix \ref{sec:csbrief}.

Random matrices, such as Gaussian matrices, effectively satisfy the incoherence and RIP conditions. The array configurations of telescopes like ASKAP (Australian Square Kilometer Array Pathfinder) and MeerKAT (Karoo Array Telescope) utilize Gaussian distribution. From the perspective of radio interferometry, the UV coverage (sampling matrix) typically cannot be assumed to follow a Gaussian distribution and may not have a closed form. However, for structured measurement matrices like UV coverage, \cite{candes_probabilistic_2011} suggest that the RIP may not be essential in practical applications; instead, incoherence often suffices. Incoherence achieves a similar effect to RIP by ensuring that each sparse element interacts with multiple measurements in a non-overlapping, distributed manner. Furthermore, \cite{wright_high_dimensional_2022} indicate that theoretical principles established for Gaussian models frequently provide accurate predictions for the behavior of \( \ell_1 \) minimization in various sampling contexts. Additionally, due to the inherent incoherence between the time and frequency domains \citep{foucart_mathematical_2013,schulz_compressive_2021}, the UV coverage of radio telescopes as a sampling matrix could meet the requirements of incoherence and, in certain cases, RIP. Consequently, the primary condition for applying CS theory in radio interferometric imaging is ensuring a sparse representation of the signal.

\subsection{From Wavelet Dictionaries to the Deep Dictionary}\label{subsec:wtdeep}

In the late 1990s, foundational concepts for most signal transformations were established, and with the advent of the Basis Pursuit method, the focus of signal representation shifted from traditional transforms to dictionaries. Classic transforms, such as wavelets, are characterized by predefined and fixed basis functions. Mathematically, wavelets are well-suited for capturing point singularities and abrupt changes within signals, making them particularly effective for representing edges and detailed structures across multiple scales. However, the inherent properties of these transforms limit their expressive capacity. For instance, the IUWT lacks directional information due to its isotropic filtering process in two-dimensional data, potentially resulting in a loss of sparsity when compared to the curvelet transform \citep{rubinstein_dictionaries_2010}. The history of dictionary development reveals several core concepts in modern dictionary design (see \cite{elad_sparse_2010,mallat_wavelet_2009,starck_sparse_2010}), four of which are crucial for dictionaries used in radio interferometry image deconvolution.

\begin{itemize}
\item  \emph{Localization}: better localization in transforms or dictionaries leads to improved sparsity because it allows the basis functions to efficiently capture the significant features of the signal.  This means fewer basis functions are needed to represent the signal accurately, leading to a sparse representation \citep{rubinstein_dictionaries_2010}. 

\item  \emph{Multi-resolution}:  Multi-resolution processing enhances spatial localization of features, enabling the capture of both fine details and global structures \citep{donoho_compressed_2006}. Since a radio interferometer suffers from the short-spacing issue, the measurement of total power and visibility functions for sources with large angular sizes is inadequate. In this context, multi-resolution analysis is crucial, as low resolution facilitates the reconstruction of large-scale structures for complex sources.

\item \emph{Overcompleteness}: in contrast to an orthogonal dictionary (where the number of basis functions equals the dimensionality of the signal space), an overcomplete dictionary has more basis functions than the signal’s dimensionality.  This redundancy provides a more flexible basis to improve sparsity \citep{rubinstein_dictionaries_2010}. In addition, the null space of an overcomplete dictionary ensures that a sparse signal can be uniquely represented \citep{elad_sparse_2010}.

\item \emph{Data-adaptive}: to achieve greater expressiveness and sparsity, one option is to adapt the transform bases to the signal content.
\end{itemize}

Building on the aforementioned four core concepts and aiming to enhance the expressive capabilities of wavelet-like dictionaries, we propose a deep dictionary with a cascaded, multi-scale architecture. The analysis step (forward transform) of our deep dictionary employs a convolutional neural network incorporating $M_s$ orthogonal filters in each layer. This approach reduces the complexity of sparse approximations selected in a redundant dictionary \citep{mallat_wavelet_2009}. The number of $M_s$ is contingent upon the dimension $K_d$ of the kernel, such that $M_s = K_d$, thereby constructing orthogonal bases in each layer. Given an input 2D signal $\mathbf{U}$ of size $ H \times W $, and a set of $ M_s$ filters $\mathbf{F} = \{\mathbf{F}_1, \mathbf{F}_2, \ldots, \mathbf{F}_{M_s}\}$   where each filter  $\mathbf{F}_m$  is of size $d \times d =K_d$, the convolution operation of each level is performed as follows:
\begin{equation}
Y = \text{ReLU}(  \mathbf{U} \ast \mathbf{F}  + B)
\end{equation}
where $Y$ is the output of the layer, and $B$ is the bias term. ReLU represents the Rectified Linear Unit operator.

\begin{figure}
\centering
\includegraphics[width=0.48\textwidth]{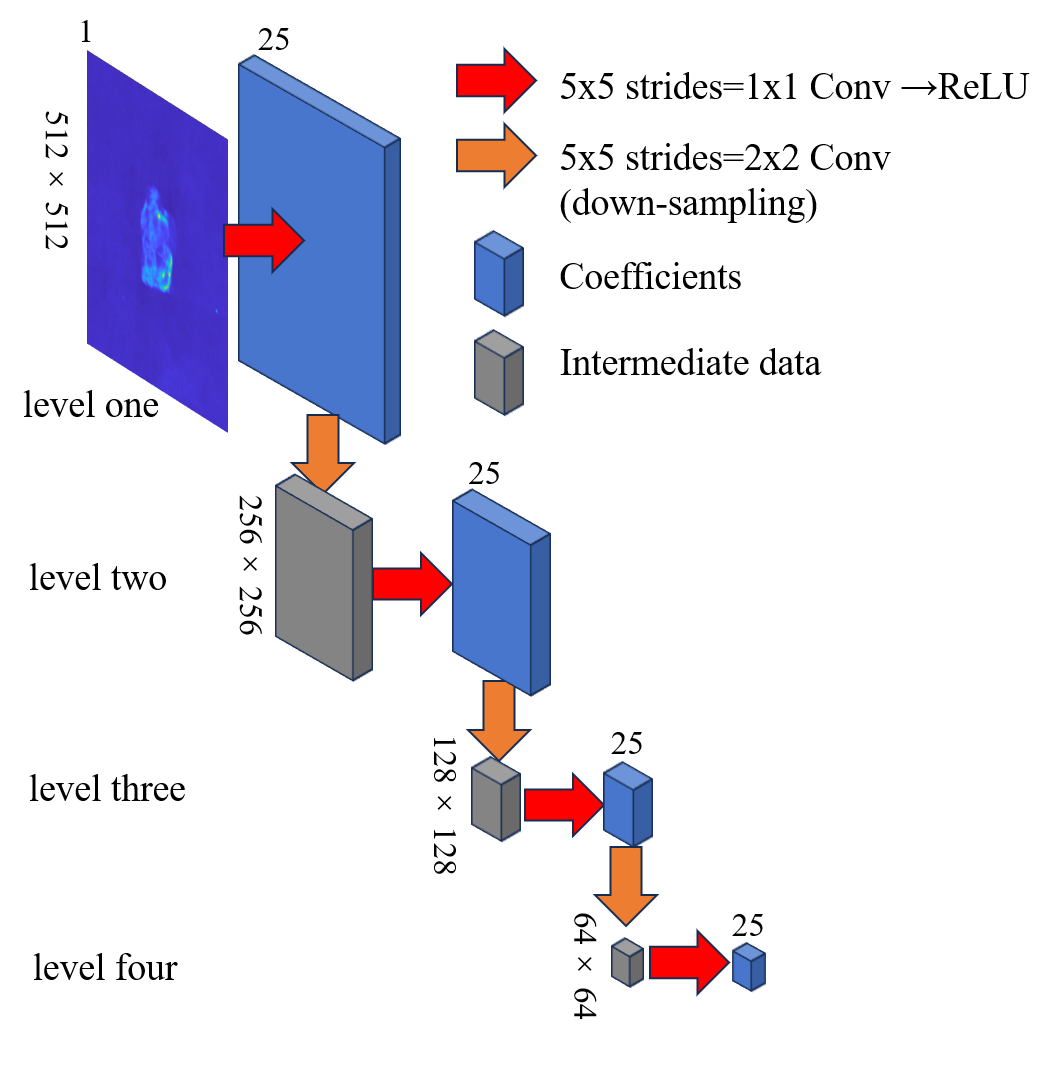}
\caption{The structure of the deep dictionary analysis step. This example features four decomposition levels, with an input signal dimension of 512x512 and orthogonal filters sized at 5x5. At each level, the signal is projected into a vector space through a convolution operation consisting of 25 bases. The output from the convolution, followed by the ReLU operator, represents the coefficients for that level. These coefficients are then down-sampled (via convolution with stride 2) and fed into the next decomposition level. Finally, the analysis process produces multi-resolution coefficients. Note that, in the example analysis step, the input channel of the first convolution is 1, while the output channel is set to 25. }
\label{fig:analysis}
\end{figure}

Since the ReLU operator is equivalent to the shrinkage (soft thresholding) operator in the non-negative case \citep{papyan_convolutional_2017}, where  $\mathscr{S}(\alpha,\beta) =\text{ReLU}(\alpha-\beta)$, the inclusion of a bias term in each convolutional layer facilitates the selection of significant coefficients and enhances sparse representation. This mechanism is equivalent to the selection of significant coefficients in the IUWT using an appropriate threshold, as demonstrated in Fig. \ref{fig:cwoef }. Notably, the threshold is adaptively learned from the input data. The structure of the dictionary analysis process is demonstrated in Fig. \ref{fig:analysis}.

\begin{figure}
\centering
\includegraphics[width=0.48\textwidth]{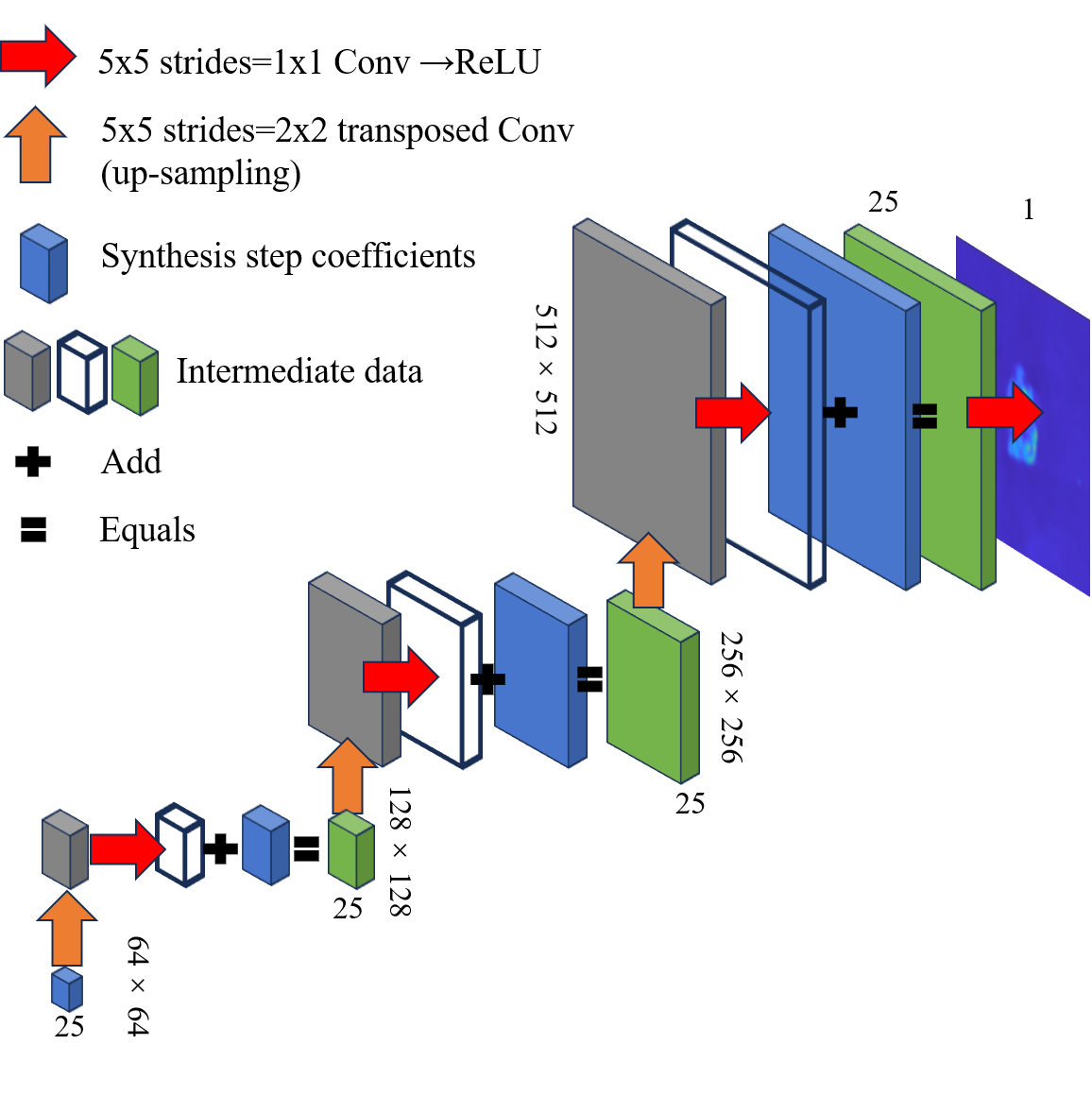}
\caption{The structure of the deep dictionary synthesis step. In each layer of the synthesis dictionary, the sparse coefficients are first up-sampled through a transposed convolution operator (with stride 2) and then convolved with orthogonal filters. This output is then added to the sparse coefficients produced from the analysis step at the same resolution. The size and number of filters in each layer are identical to those in the analysis step. Note that, in the example synthesis step, the input channel of the last convolution is 25, while the output channel is set to 1. }
\label{fig:synstep}
\end{figure}

The synthesis step (inverse transform) of the deep dictionary is also a convolutional neural network. It reconstructs a signal from the sparse coefficients generated during the analysis step, sequentially utilizing these coefficients from low to high resolution. Figure \ref{fig:synstep} illustrates the structure of the synthesis step. 

The deep dictionary is overcomplete and multi-resolution, with its basis vectors learned from data. While the bases within each level of the dictionary remain orthogonal, the bases across different levels are non-orthogonal. 
\section{The proposed method\label{sec:proposedm}}

Appendix \ref{sec:radioi} provides a brief overview of the theory of radio interferometric imaging. Based on the principle of radio interferometry and CS, we assume that the signal can be expressed in a sparse form. Then substitute the variables from equations (\ref{eq:vm}) and (\ref{eq:ny}) into equation (\ref{eq:resnoise}), the CS-based deconvolution method can be formulated as follows:
\begin{equation}
\min_I \; \|\Psi I\|_1 \;\text{subject to} \;\|M\mathcal{F} I - V_m\|_2^2 \leq \epsilon
\label{eq:csdeconv}
\end{equation}
where $I$ denotes the sky brightness distribution, $\Psi$ represents a dictionary (such as wavelet dictionary), $M$ is the $(u, v)$ coverage, and $V_m$ corresponds to the measurement visibility.

To preserve spatial structure and balance frequency recovery with noise suppression, we incorporated TV as an additional regularization. TV regularization in deep neural networks can improve generalization and mitigate overfitting. From the perspective of interferometric deconvolution, TV helps estimate image sparsity in the gradient domain, significantly influencing the effective spatial resolution of the reconstructed image. Following the recommendations of \cite{getreuer_total_2012} and \cite{akiyama_superresolution_2017}, we incorporate TV into Equation (\ref{eq:csdeconv}), resulting in the following mathematical model for signal recovery:

\begin{equation}
\min_I \; \|\nabla I\|_1+\|\Psi I\|_1 \;\text{subject to} \;\|M\mathcal{F} I - V_m\|_2^2 \leq \epsilon
\label{eq:bvn}
\end{equation}
Where $||\nabla I||_1 $ is the TV regularization term. $\Psi$ means a transform or dictionary. In the anisotropic scenario, which treats the horizontal $x$ and vertical $y$ directions separately, the unconstrained form of equation (\ref{eq:bvn}) becomes:
\begin{equation}
\min_I \; \| \nabla_x I\|_1+ \| \nabla_y I\|_1+\|\Psi I\|_1 +\frac{\mu}{2}||M\mathcal{F} I - V_m||_2^2 
\label{eq:spbv}
\end{equation}
Where $\mu$ is a parameter that controls the contribution from the $\ell_2$-norm part.

\subsection{Deconvolution  with a Fixed Dictionary}

Initially, we assume that the dictionary is fixed, as in the case of wavelet bases. To address the optimization problem outlined in equation (\ref{eq:spbv}), we employed the split Bregman method \citep{goldstein_split_2009}, which is particularly well-suited for $\ell_1$-regularized problems.The split Bregman method efficiently solves $\ell_1$-regularized optimization problems by introducing auxiliary variables, using a combination of first-order gradient descent and shrinkage, and iteratively updating the variables until convergence. This method is capable of handling multiple $\ell_1$-norm terms and exhibits faster convergence compared to widely used proximal gradient methods, such as ISTA and FISTA, which often require more careful tuning of step sizes. While the underlying mathematical framework of the split Bregman method is close to the alternating direction method of multipliers (ADMM) \citep{boyd_distributed_2010}, it specifically exploits Bregman iteration to handle non-smooth terms, making it a preferred approach for sparsity-promoting applications. In order to apply the split Bregman method, which decouples the $\ell_1$ and $\ell_2$ components, it is necessary to introduce auxiliary variables. Therefore, the constrained formulation of the problem becomes:
\begin{equation}
\begin{split}
\min_{I,d_x,d_y,w} \|d_x\|_1+\|d_y\|_1 +\|w\|_1+\frac{\mu}{2}\|M\mathcal{F}I-V_m\|_2^2  \\
\text{such to} \quad d_x=\nabla_xI,\;d_y=\nabla_yI,\;w=\Psi I
\end{split}
\label{eq:consb}
\end{equation}
Where $d_x$, $d_y$ and $w$ are auxiliary variables. According to the split Bregman method \citep{goldstein_split_2009}, the associated unconstrained problem to equation (\ref{eq:consb}) is defined as:
\begin{equation}
\begin{split}
\min_{I,d_x,d_y,w} \|d_x\|_1+\|d_y\|_1 + \|w\|_1+\frac{\mu}{2}\|M\mathcal{F}I-V_m\|_2^2 \\
+\frac{\lambda}{2}\|d_x-\nabla_x I -b_x\|_2^2+\frac{\lambda}{2}\|d_y-\nabla_y I -b_y\|_2^2 \\
+\frac{\gamma}{2}\|w-\Psi I -b_w\|_2^2
\end{split}
\label{eq:consbv}
\end{equation}
Where $\mu$, $\lambda$ and $\gamma$ are positive parameters that balance the components of the equation. The variables $b_x$, $b_y$ and $b_w$ are associated with the Bregman iteration algorithm. For the solution details see Appendix \ref{sec:ssbm}.

The parameters $ \mu $, $ \lambda $, and $ \gamma $ in equation (\ref{eq:consbv}) balance the contributions of regularization terms. To optimize these parameters, we enhance the Bregman iteration algorithm by incorporating an unsupervised deep learning approach. Notably, the max operator in the split Bregman method can be seamlessly substituted with the ReLU operator, which in its reformulated state is equivalent to the shrinkage operator. Therefore, the split Bregman method can be reconstructed as a neural network. Further details on this reformulation are provided in Appendix \ref{sec:refsbm}.

To preserve the core principles of the split Bregman framework while enhancing the adaptability of the network, we allow $\mu$, $\lambda$, and $\gamma$ to vary dynamically across iterations. The resulting loss function is defined as:
\begin{equation}
 \mathcal {L}_{\text{fidelity}} =\|M\mathcal{F}I-V_m\|_2^2
\label{eq:fixd_deconv}
\end{equation}
The complete process for this method is outlined in Algorithm \ref{alg:fix}. 

\begin{algorithm}[H]
	\caption{Deconvolution  with Fixed Dictionaries }
	\begin{algorithmic}
        \State \textbf{Initialize:} {$I^0=\mathcal{F}^{\dagger} V_m$}
        \State \textbf{Initialize:} $d_x^0=d_y^0=w^0=b_x^0=b_y^0=b_w^0 =0$
        \State \textbf{Initialize:} $\mu^0$,$\lambda^0$ and $\gamma^0$, $K^0=\mu^0 M^TM-\lambda^0\Delta+\gamma^0 \mathbf{I}$

         \While {$||M\mathcal{F}I -V_m ||^2_2 > \epsilon_d $}
            \For {$i$=1 to N}
            \State Execute  equations (\ref{eq:s0}) and (\ref{eq:s1}), (\ref{eq:ns2}) to (\ref{eq:ns4}), (\ref{eq:s5}) to (\ref{eq:s8}) in sequence
            \EndFor
            \State Update $\mu$,$\lambda$ and $\gamma$
            
        \EndWhile
        \State Note that $\epsilon_d$ is a predefined value that represents the uncertainty in the $V_m$.
	\end{algorithmic}
 \label{alg:fix}
\end{algorithm}


\subsection{Deconvolution with a Deep Dictionary}

In order to improve the expression capability of predefined bases dictionaries, we applied our deep dictionary to Algorithm 1. In the context of loss function design, it should primarily include the fidelity constraint, represented as $\|M\mathcal{F}I-V_m\|_2^2$. Additionally, given that the dictionary employed here is a deep dictionary, it must satisfy the symmetry constraint $\Psi^T(\Psi(V_m)) = V_m$. Moreover, the base vectors must satisfy orthogonality within each layer. According to the findings by \cite{kim_revisiting_2022}, kernel orthogonality regularization (KOR) outperforms other methods. Here, we apply an efficient algorithm for KOR called the spectral restricted isometry property (SRIP). Therefore, the loss function is as follows:

\begin{equation}
\mathcal {L}_{\text{total}} = \mathcal {L}_{\text{fidelity}} + \alpha_1\mathcal {L}_{\text{symmetry}}+
\alpha_2\mathcal {L}_{\text{orthogonality}}\\
\end{equation}
with:
\begin{eqnarray}
\mathcal {L}_{\text{symmetry}} = \|\Psi^T(\Psi(V_m))-V_m\|_2^2 \label{wq:sym_loss} \\
\mathcal {L}_{\text{orthogonality}} = \sigma(\mathbf{F}^T\mathbf{F}-\mathbf{I}) \label{eq:or_loss}
\end{eqnarray}

where $\mathbf{F} \in \mathbb{R}^{d\times d} $ represents the basis vectors (filters), $\alpha_1$ and $\alpha_2$ are positive values that control the contribution from the constraints. The symbol $\sigma(.)$ denotes the spectral norm. The spectral norm of a matrix $A\in \mathbb{R}^{m\times n}$ can be expressed as \citep{yoshida_spectral_2017}:
\begin{equation}
\sigma(A) = \max_{\xi\in \mathbb{R}^n \,,\xi \ne 0}\frac{||A\xi||_2}{||\xi||_2}
\end{equation}
where $\xi \in \mathbb{R}^n $ represents a perturbation vector with a small $\ell_2$-norm.


The proposed unsupervised learning with the deep dictionary, the Deep Split
Bregman Network (DSB-net), is presented in Algorithm \ref{alg:deepl_deconv}.
\begin{algorithm}[H]
	\caption{ Deconvolution with the Deep Dictionary}
	\begin{algorithmic}
        \State \textbf{Initialize:} {$I^0=\mathcal{F}^{\dagger} V_m$}
        \State \textbf{Initialize:} \text{The deep dictionary decomposition level}
        \State \textbf{Initialize:} \text{Filter size and the number of filters} $M_s$ 
        \State \textbf{Initialize:} $d_x^0=d_y^0=w^0=b_x^0=b_y^0=b_w^0 =0$
        \State \textbf{Initialize:} $\mu^0$,$\lambda^0$ and $\gamma^0$, $K^0=\mu^0 M^TM-\lambda^0\Delta+\gamma^0 \mathbf{I}$
        \State \textbf{Initialize:} $\alpha_1$,\,$\alpha_2$
        \State \textbf{Initialize:} optimizer and Learning rate  

         \While {$\mathcal {L}_{\text{total}}$ and $\mathcal {L}_{\text{fidelity}}$ are not converged}
            \State $\textbf{given the dictionary, update the image}$:
            \State  \hspace{\algorithmicindent}  Execute  equations (\ref{eq:s0}) and (\ref{eq:s1}), (\ref{eq:ns2}) to (\ref{eq:ns4}), (\ref{eq:s5}) to (\ref{eq:s8}) in sequence

            \State $\textbf{given the image, update the dictionary}$:
            \State \hspace{\algorithmicindent} Update the dictionary $\Psi$ and other parameters 
            
        \EndWhile
	\end{algorithmic}
 \label{alg:deepl_deconv}
\end{algorithm}

\subsection{Training Details} \label{traind}

Throughout the training process, the recovered image and the deep dictionary (or parameters) are alternately updated. Given a dictionary, the recovered image is computed using the split Bregman method, after which the deep dictionary (or parameters) is updated. For the proposed method utilizing the DWT dictionary, the parameters $\mu^0 $, $\lambda^0 $, and $\gamma^0 $ are set to 40, $1/3N_{\sigma}$, and $1/N_{\sigma} $, respectively. The decomposition level is set to 5, consistent with the IUWT-based CS method(see the section \ref{sec:experimentalr} experimental results), and the maximum number of iterations is 1000. The Adam\footnote{ \url{https://pytorch.org/docs/stable/generated/torch.optim.Adam.html}} optimizer is employed with a learning rate of  $0.1$. The convergence condition $\mathcal{L}_{\text{fidelity}}$ is set to $1 \times 10^{-4}$.

For the proposed method employing the deep dictionary, the initial values of $\mu^0$, $\lambda^0$, and $\gamma^0$ are set as $40$, $100$, and  $1/3N_{\sigma}$, respectively.  To accelerate convergence, the deep dictionary is held fixed during the first 100 iterations, while $\mu_0$, $\lambda_0$, and $\gamma_0$ are updated iteratively. The decomposition level is set to 5, and the maximum number of iterations remains 1000. The Adam optimizer is used with a learning rate of $0.001$.

The filter size (base vector length) is determined by the field of view, corresponding to the synthesized beam size (see the section \ref{sec:experimentalr} experimental results); hence, it is set to $5\times5$, slightly larger than the synthesized beam. The number of base vectors (filter channels) per level is equal to the filter dimensions ($25$ for a $5\times5$ filter). The parameter $\alpha_1$ controls the completeness of the reconstruction. Since the method recovers the input signal from significant analysis coefficients and does not require perfect reconstruction (with the difference tending to be noise), $\alpha_1$ is set to $0.001$. The orthogonality term $\alpha_2$ has a strong effect in enforcing filter orthogonality, making its choice relatively insensitive \citep{kim_revisiting_2022}; thus, $\alpha_2$ is set to $0.01$. Finally, the convergence conditions $\mathcal{L}_{\text{total}}$ and $\mathcal{L}_{\text{fidelity}}$ are set to $0.7$ and $1 \times 10^{-5}$, respectively.

\section{Experimental results\label{sec:experimentalr}}

In this section, we simulate observations from the Karl G. Jansky Very Large Array (VLA) telescope using \textsf{CASA}\footnote{\url{ https://casa.nrao.edu/}} (Common Astronomy Software Applications) to verify the performance of the proposed method. The B configuration of the VLA is employed in the simulation. To evaluate the performance of our proposed algorithms, we will compare them with existing deconvolution algorithms. The evaluation will encompass our two approaches: one employing a fixed dictionary based on the discrete wavelet transform (DWT) and the other utilizing a deep dictionary. This comparative analysis facilitates the estimation of performance improvements attributable to the deep dictionary. The results will be compared to those obtained by multiscale CLEAN (MS-CLEAN), the deconvolution method developed by \cite{li_application_2011} (IUWT-based CS method), and the synthesis-analysis dictionary reconstruction method MORESANE \citep{dabbech_moresane:_2015}. Note that the \textsf{CASA tclean} function will be used to implement the MS-CLEAN deconvolution. Assuming the sky brightness is the emission of a specific spectrum line, the simulated observation includes one channel at 1.4\,GHz with a bandwidth of 0.1\,MHz and an integration time of 60 seconds. The corresponding synthesized beamwidth is approximately 4.3 arcseconds. Robust weighting is adopted, setting the briggs robustness parameter to 0. The pixel size is set to 1 arcsecond, which corresponds to approximately one-fifth of the angular resolution of the VLA telescope. The simulated radio sources are positioned at right ascension $\mathrm{12^h42^m44.3^s}$ and declination $\mathrm{32^d16^m9^s}$, and the VLA array is located at a latitude of approximately 34 degrees.

\begin{figure}
\centering
\includegraphics[width=0.48\textwidth]{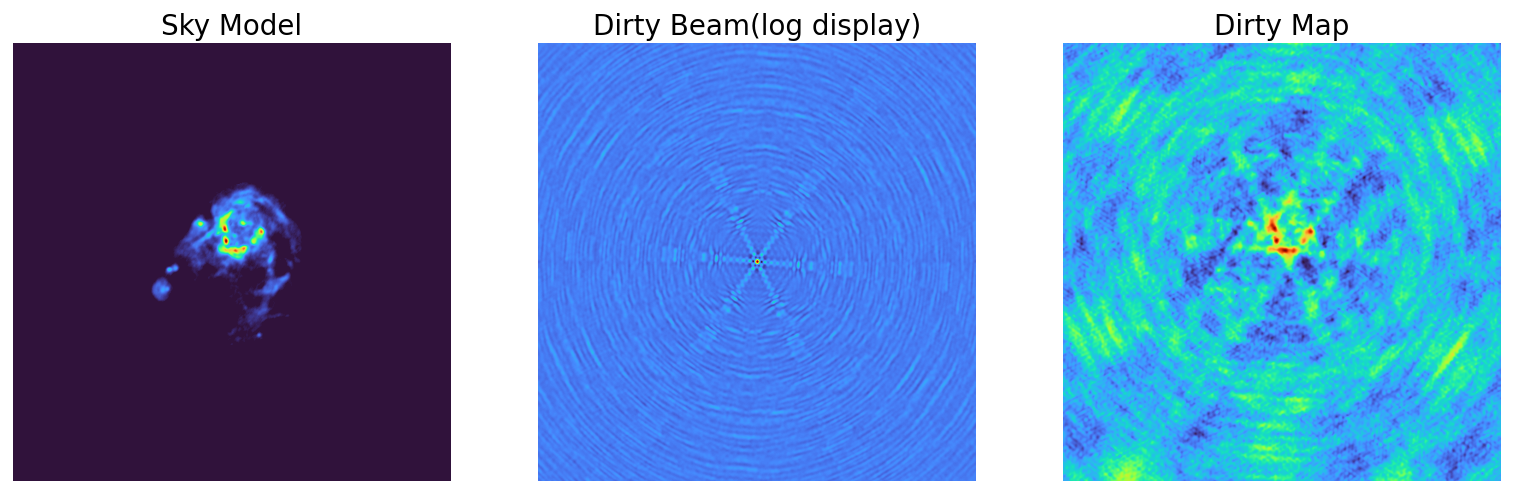}
\caption{The simulated observation of an HII region in M31. \emph{Left}: the sky model, where the brightness ranges from $1\times 10^{-9}$ to $1.006$ Jy/pixel. \emph{Middle}: the dirty beam displayed with logarithmic scaling. \emph{Right}: the observed dirty map. All images demonstrated the central regions.}
\label{fig:m31-obs}
\end{figure}

The first test sky model image, shown on the left of Fig. \ref{fig:m31-obs}, is an HII region in M31 that has been extensively used in previous studies (such as \cite{li_application_2011,dabbech_moresane:_2015}).  Noise in the simulated visibility data was generated using the \textsf{CASA simulator.setnoise}\footnote{Available at: \url{https://casadocs.readthedocs.io/en/stable/api/tt/casatools.simulator.html\#casatools.simulator.simulator.setnoise}} function, resulting in a noise standard deviation ($N_{\sigma}$) of approximately $0.1218$ Jy/pixel. And the integrated signal-to-noise ratio (SNR) is 11.88. Given that the primary beam size of the VLA at 1.4\,GHz is approximately $30$ arcminutes\footnote{\url{ https://science.nrao.edu/facilities/vla/docs/manuals/oss/performance/fov}}, the simulated image size was set to $1024\times 1024$ pixels. The sky model image (the original size is $256\times 256$) was padded with zeros to fit the simulation. In this simulated observation, a snapshot mode was considered, with a total observation time of 1 hour. Consequently, the synthesized beam contains high sidelobes, as shown in the middle of Fig. \ref{fig:m31-obs} (displayed in logarithmic scale). The dirty image, corrupted by the point spread function (PSF) or dirty beam, is shown on the right of Fig. \ref{fig:m31-obs}.

\begin{figure*}
\begin{center}
\makebox[\textwidth]{\includegraphics[width=0.7\paperwidth]{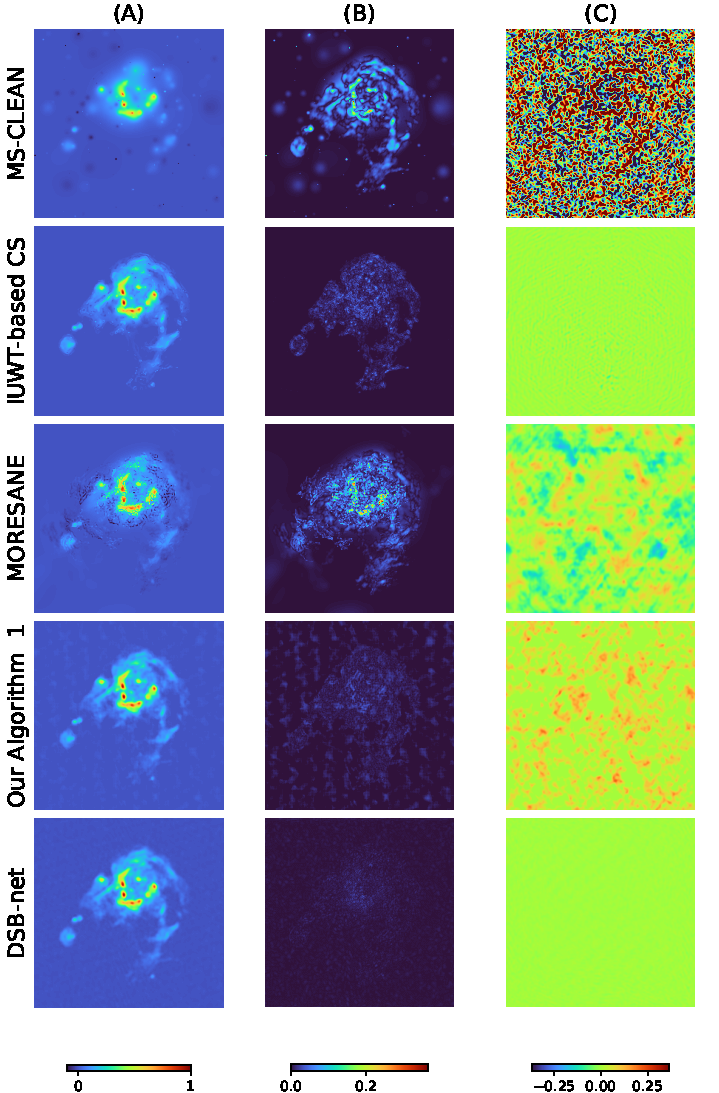}}
\end{center}
\caption{Reconstructed images of the simulated M31 observation with VLA. \emph{from top to bottom} in rows are the results of the MS-CLEAN, the IUWT-based CS, the MORESANE method, the proposed method employing the DWT dictionary, and the proposed DSB-net, respectively. \emph{From left to right}, $\mathbf{(a)}$ the model images, $\mathbf{(b)}$ model absolute error images (the absolute differences between the reference model images and the reconstructed model images), and $\mathbf{(c)}$ the residual images. The color bar range for the error images is set from 0 to  $3N_{\sigma}$, while for the residual images, it is configured from $-3N_{\sigma}$ to $3N_{\sigma}$}
\label{fig:m31-compare}
\end{figure*}

\begin{figure*}
\centering
\includegraphics[width=0.8\paperwidth]{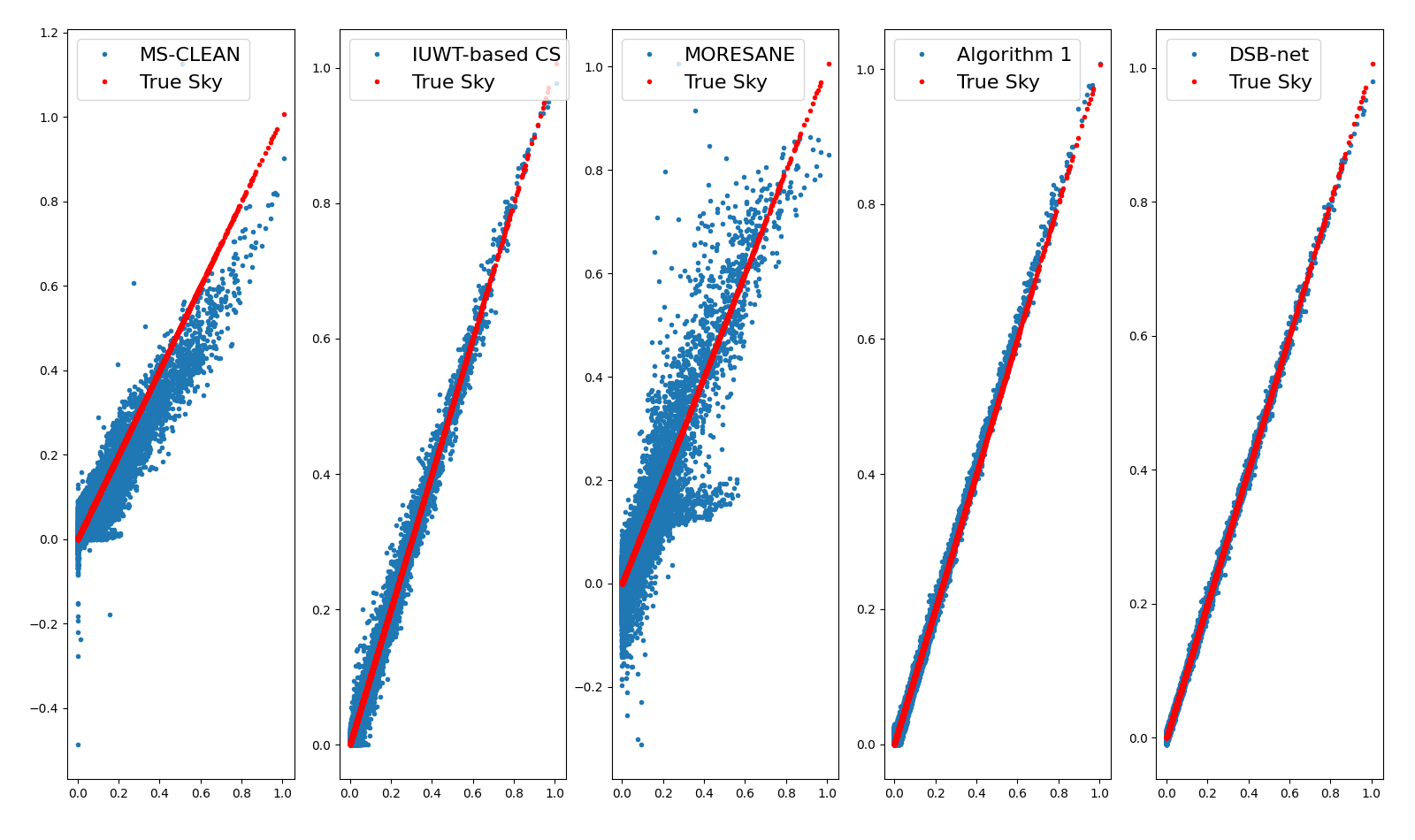}
\caption{\emph{From left to right}: The recovery results of the M31 sky model using the MS-CLEAN, IUWT-based CS, MORESANE, and the proposed method, respectively. The models are plotted along the y-axis and the true sky image is plotted along the x-axis(the blue dots). The red dots represent the true sky image values.}
\label{fig:m31-compareTrue}
\end{figure*}

The configurations of the MS-CLEAN method, the IUWT-based CS method, the MORESANE method, and our proposed methods are detailed in Appendix \ref{sec:m31mconfig}.

The deconvolution results are illustrated in Fig. \ref{fig:m31-compare}. Columns A, B, and C display the model images, model absolute error images (the absolute differences between the reference model images and the reconstructed model images), and the residual images, respectively. From top to bottom, the methods shown are MS-CLEAN, MORESANE, the IUWT-based CS, our method with DWT, and our DSB-net. The qualitative comparison indicates that the performance of DSB-net surpasses that of MS-CLEAN, MORESANE, and the IUWT-based CS method, as well as the proposed approach utilizing the DWT dictionary. This superior performance can be attributed to the incorporation of TV regularization and the expressive capability of the deep dictionary. It should be noted that all displayed images are focused on their central regions.

A pixel-by-pixel photometric comparison is shown in Fig. \ref{fig:m31-compareTrue}. In the reconstructions of M31, our recovery image(from the DSB-net) is much closer to the true sky image and demonstrates better noise suppression compared to the IUWT-based CS method. 
\begin{table*}
   \caption{The numerical comparison of the M31 deconvolution results is presented in terms of root mean square error and dynamic range. Note that robust weighting is applied to the UV coverage.} 
   \label{tab:m31ncom}
    \centering
    \begin{tabular}{ c c c c c c}
\hline
  &MS-CLEAN&IUWT-based CS&MORESANE&  Algorithm 1 &DSB-net\\
\hline

RMS&0.3649 & 0.0121& 0.0532& 0.0602 & 0.0049\\
DR& 48 & 870 & 206 & 152 & 2173\\
\hline
    \end{tabular}
\end{table*}

The concepts of root mean square error(RMS) and dynamic range(DR) are utilized for numerical comparison \citep{li_application_2011}. The RMS error is defined as:

\begin{equation}
\text{rms error}=\sqrt{  \frac{\sum(\text{residual image})^2}  {\text{number of pixels}}    }
\end{equation}
and the DR is formulated  as:
\begin{equation}
\text{DR}=\frac{\max(\text{restored image})} {\text{rms error}}
\end{equation}
where the \textit{restored image = clean beam $\ast$ model + residual image}.
Table \ref{tab:m31ncom} presents the results, demonstrating that DSB-net achieves the highest reconstruction quality in terms of both RMS and DR. Specifically, DSB-net’s RMS error is approximately 1\% of that achieved by MS-CLEAN, while its DR is about 45 times higher than that of MS-CLEAN.

The second experiment took the 3C\,288 radio galaxy\footnote{ \url{https://casaguides.nrao.edu/index.php/Sim_Inputs}}
as the sky model, chosen for its large-scale structure. Details of the simulated observation, methodological configurations, and comparative results with other methods are presented in Appendix \ref{sec:3c288}. The findings are consistent with those obtained for the M31 model. From a theoretical perspective, our approach is designed for reconstructing extended sources. However, to facilitate a comparative analysis of different methods, we also employ a sky model that includes prominent point sources. The details of this simulation are provided in Appendix \ref{sec:pskym}. Additionally, Appendix \ref{sec:msinflu} examines how the multi-resolution property of the dictionary influences the reconstruction of large-scale structures.


\section{Discussion and conclusion\label{sec:conclusion}}

In practical applications of radio interferometry deconvolution, the proposed method effectively balances the recovered and true visibility data under noisy conditions, achieving an optimal trade-off between sparsity and error minimization. This equilibrium hinges on the intricate relationship between overfitting and the generalization capacity of the network. 



For the majority of M31's brightness values, which fall below the noise level, DSB-net demonstrates a remarkable capability to recover these components, outperforming traditional, non-learning-based methods, such as wavelet dictionary approaches. The deep neural network's generalization ability is pivotal in this regard, as it enables the model to extract the underlying signal from the noisy data.  In contrast, when a network learns not only the underlying data structure but also the noise specific to the training data, heavy overfitting can occur, potentially degrading performance.

In deep learning methods, a common objective is to uncover the underlying data structure, which may involve learning a mapping from noisy inputs to their corresponding clean outputs. The $\ell_1$-norm regularization promotes sparsity in solutions, encouraging the model to minimize the number of non-zero coefficients in its data representation. This property is particularly relevant for inverse problems, as it enables the model to identify the key features necessary for accurate reconstruction while simultaneously disregarding noise or irrelevant details \citep{hastie_elements_2009}. This mechanism mitigates overfitting by using only a limited number of significant coefficients to represent the signal. This approach is analogous to polynomial regression, which penalizes large coefficients and promotes smaller weights, thereby reducing model complexity and minimizing overfitting. Based on this discussion, we investigate the overfitting prevention capability of our deep dictionary. The decomposition level of the deep dictionary is set to 5, with a base vector size of 25. To ensure overcompleteness, the minimum number of base vectors per level is set to 6. Thus, each dictionary level is set to contain 6 base vectors and retain other configurations of DSB-net. There are no significant differences appeared when compared to the parameter set at 25 (e.g., RMS error for M31 is 0.0080). These findings indicate that the proposed method effectively prevents overfitting.


In the context of radio interferometer deconvolution, we underscore the significance of the CS methodology. The theory provides a robust framework for the stable and unique recovery of sky brightness in noisy environments when specific conditions are met. For many radio interferometric telescopes, these conditions can be simplified to a sparse representation. We then outlined four key concepts for dictionary design and developed a deep dictionary to enhance the sparse representation capability for extended sources. Furthermore, we integrated the dictionary within the CS mathematical framework, resulting in a fully interpretable unsupervised learning method.

We conducted two experiments to evaluate the performance of DSB-net, and the results demonstrate that the proposed method is highly competitive with both CS-based and CLEAN-based techniques in complex morphologies. Notably, the RMS error of DSB-net is approximately 1\% of that achieved by MS-CLEAN, while its DR is nearly 45 to 100 times higher than that of MS-CLEAN. As an unsupervised learning approach, DSB-net does not rely on extensive training datasets, which significantly reduces associated costs and enhances robustness. Additionally, our method bridges the gap between deep learning and classical dictionary techniques, such as wavelets.

However, the presence of sources below the noise level (3$\sigma$) complicates the determination of the convergence threshold and makes it challenging to identify unknown sources. In practice, we can set the maximum number of iterations or the lower bound of $1/\lambda$ and $1/\gamma$ to control this balance. 

We evaluate the execution time of our proposed methods with that of MS-CLEAN by running each program 100 times on the same data. The runtime of our method with a fixed dictionary is shorter than that of MS-CLEAN, whereas the method incorporating a deep dictionary requires more time due to the additional computational cost associated with learning the basis. Our methods are executed on a single RTX 4090 GPU, while MS-CLEAN is executed on a CPU. The comparison is conducted across different image sizes, and the reported times represent the mean run-times with their corresponding standard deviations, as presented in Table \ref{tab:timeC}.
The proposed method is implemented in Python 3.12, utilizing PyTorch 2.21 and CUDA 12.1. The computer specifications are listed in Table \ref{tab:comc}. The code is available at \url{https://github.com/MoerAttempts/the-Deep-Split-Bregman-Deconvolution-Network}.

\begin{table*}
   \caption{Comparison of run-times to the proposed methods and MS-CEALN for different image sizes. \label{tab:timeC}}
    \centering
    \begin{tabular}{ c c c c}
\hline
  Image size [pixels]&MS-CLEAN run-time [s]& Algorithm 1 run-time [s]& DSB-net run-time [s]\\
\hline

512&223.96 $\pm$ 0.18 & 8.83 $\pm$ 0.59& 1204.03 $\pm$ 0.18\\
1024& 288.73 $\pm$  0.10 & 22.17 $\pm$ 0.19& 4743.18 $\pm$ 1.02\\
\hline
    \end{tabular}
\end{table*}

\begin{acknowledgements}
We acknowledge partial support from the Science and Technology Innovation Program of Hunan Province under grant number 2024JC0001. This research can serve as a theoretical exploration of FASTA(FAST Array) imaging methods.
\end{acknowledgements}


\begin{appendices}

\section{An Overview of Previous Deep Learning-based Studies}\label{sec:dlm}

Previous studies mainly employ the learning power of deep learning networks. For example, \cite{nammour_shapenet_2022} introduced a deep learning method named ShapeNet, designed to enhance the deconvolution of galaxy images in both optical astronomy and radio interferometry by incorporating shape constraints. This method extends the Tikhonet method by incorporating a shape constraint, thereby better preserving the shape of galaxies and reducing pixel error.
\cite{schmidt_deep_2022} employ  convolutional neural networks (CNNs) inspired by super-resolution models to achieve reconstruction based on noise-simulated data. They train models on simulated data of radio galaxies composed of Gaussian components. \cite{chiche_deep_2023} propose a deep learning-based solution for signal reconstructions across various noise levels in radio astronomy. Their method focuses on spatial and temporal deconvolution, capturing the temporal structure of the sky. Additionally, \cite{geyer_deep-learning-based_2023} present a technique for simulating images using generative adversarial networks (GANs) to reconstruct missing visibility data from sparse radio interferometer observations. This approach is an improved version of their previous work, which relied on CNNs.
\cite{terris_image_2022} propose iterative image reconstruction algorithms that combine convex optimization and deep learning. Their approach involves training a deep neural network as a denoiser, with the iterative process alternating between denoising and gradient-descent data-fidelity steps. QuantifAI, introduced by \cite{liaudat_scalable_2024}, offers a scalable and efficient approach to Bayesian uncertainty quantification (UQ) in radio interferometric imaging. By leveraging data-driven priors and convex optimization techniques, the method achieves high-quality reconstructions and meaningful UQ. The prior potential is constructed by training a neural network as a denoiser.



\section{Radio Interferometry\label{sec:radioi}}

In general scenarios, an observer at location $\boldsymbol{r}$  receives electromagnetic fields from a distant astronomical object, which can be expressed as $E_\mathit{v}(\boldsymbol{r})$.   This radiation exhibits spatial incoherence, we consider it as a scalar and quasi-monochromatic quantity. An interferometer measures the spatial coherence function of the electromagnetic field using a pair of antennas, at two different positions $\boldsymbol{r_1}$ and $\boldsymbol{r_2}$. Each measure yields complex visibility:
\begin{equation}\label{eq:bv}
      V_{\mathit{v}}(\boldsymbol{r_1},\boldsymbol{r_2})=\langle\,E_{\mathit{v}}(\boldsymbol{r_1})E_{\mathit{v}}^\ast(\boldsymbol{r_2})\rangle
\end{equation}
where $^\ast$ is  the complex conjugate and $\langle\,\rangle$ means the time average operator.

The spatial coherence function can be approximately expressed as the following, which only depends on the baseline vector $\boldsymbol{r_1}-\boldsymbol{r_2}$ \citep{taylor_synthesis_1999}.
\begin{equation}\label{eq:appv}
V_{\mathit{v}}(\boldsymbol{r_1},\boldsymbol{r_2})=\int I_v(\boldsymbol{s}) e^{-2\pi i v \boldsymbol{s}(\boldsymbol{r_1}-\boldsymbol{r_2})/c} \,d\mathrm{\Omega}
\end{equation}
where $I_v(\boldsymbol{s})$ represents the sky brightness distribution in the direction $\boldsymbol{s}$, and $\boldsymbol{s}$ is the unit vector. $c$ is the speed of light, and $d\mathrm{\Omega}$ is the solid angle.

Conventionally, the baseline vector $\boldsymbol{r_1}-\boldsymbol{r_2}$ in Eq. \ref{eq:appv} measures in terms of the unit of wavelength, it has $(u,v,w)$ components in a right-handed coordinate system, where the component $w$ is measured in the direction of the phase reference position. And $(u,v)$ components define a plane perpendicular to the $w$. The origin of source intensity distribution $I(l,m)$ lies in the phase reference position, where $(l,m)$ are direction cosines measured with respect to the $(u,v)$ axis. We assume that the antenna response is isotropic, then Eq. \ref{eq:appv} can be rewritten as \citep{thompson_interferometry_2017}:
\begin{equation}
V(u,v,w)=\iint  I(l,m) e^{-2\pi i (ul+vm+w(\sqrt{1-l^2-m^2}-1)}\frac{dl \, dm}{\sqrt{1-l^2-m^2}}
\end{equation}
The term $2\pi w(\sqrt{1-l^2-m^2}-1) $ can be neglected when its magnitude is significantly less than the unity.  Consequently, the relationship between the spatial coherence function $V(u,v)$ and the intensity distribution $I(l,m)$ becomes a two-dimensional Fourier transform.
In practice, a radio telescope array measures partial Fourier coefficients of the sky brightness, which is the $(u, v)$ coverage of the observations. Therefore, after the necessary calibration step on the visibilities, the measured visibilities $V_m$ correspond to incomplete Fourier coefficients. This situation can be described as:
\begin{equation}
V_m = M \cdot (V+\epsilon) 
\label{eq:vm}
\end{equation}
where $M$ is the $(u, v)$ coverage and $\epsilon$ is the white Gaussian noise. 

In the image domain, the inverse Fourier transform of $V_m$ and  $(u,v)$ coverage is called dirty image and dirty beam (or point spread function, PSF) respectively. The dirty image arises from the convolution of the sky brightness distribution $I(l,m)$ with the dirty beam. Hence, the measured visibilities in the image domain can be expressed as:
\begin{equation}
y = \mathcal{F}^{\dagger} M \mathcal{F}I + \mathcal{F}^{\dagger}M\epsilon
\label{eq:ny}
\end{equation}
where $y$ is the dirty image, $ I \in \mathbb{R}^N$ is the image of sky brightness, $\mathcal{F}$ and $\mathcal{F}^{\dagger}$ represent Fourier and inverse Fourier transform.


\section{A Brief Introduction to Compressive Sensing Theory}\label{sec:csbrief}

Within the conventional framework of CS theory, the fundamental objective is to recover an unknown $k$-sparse signal $u\in \mathbb{C}^n$ from an underdetermined linear system $f=\Phi u$. Here, the matrix $\Phi \in\mathbb{C}^{m\times n}$ represents the measurement matrix, with $m \ll n$ and $f\in \mathbb{C}^m$. The signal $u$ exhibits $k$ non-zero components within the sparse domain $\Psi$, where $k<n$.  This problem can be formulated as \citep{vidyasagar_introduction_2020}:
\begin{equation}
\min_u \; \|\Psi u\|_0 \;\text{subject to} \;\Phi u = f
\end{equation}
where $\|.\|_0$ represents the $\ell_0$ pseudo-norm.

The $\ell_0$-minimization problem is combinatorial and NP-hard, requiring an exhaustive brute-force search to obtain the absolute minimum.  An alternative approach(a convex relaxation of $\ell_0$-norm), the $\ell_1$-minimization  provides a fast and equivalent solution \citep{donoho_for_2006}. Then the recovery algorithm becomes:
\begin{equation}
\min_u \; \|\Psi u\|_1 \;\text{subject to} \;\Phi u = f
\end{equation}
where $\|.\|_1$ represents the $\ell_1$-norm.

Since the $\ell_1$-norm is a convex problem, it can be solved efficiently by convex optimization methods. In most practical situations, the measurements are noisy and the signal is not exactly sparse. Consequently,  $k$-sparse representation refers to the $k$ largest components of the signal (an approximate sparse signal). The reconstruction can be expressed as:
\begin{equation}
\min_u \; \|\Psi u\|_1 \;\text{subject to} \;\|\Phi u - f\|_2^2 \leq \epsilon
\label{eq:resnoise}
\end{equation}
where $\epsilon$ describes the amount of uncertainty in the data and $||.||_2$ represents the $\ell_2$-norm.

The conditions of sparsity, incoherence, and the restricted isometry property (RIP) collectively ensure stable and unique signal recovery in the presence of noise.

Sparsity implies that most elements in the signal $ u $, or its representation in a transformed domain  $\Psi u$, are zero or can be approximated with only a few non-zero elements, with $ k $ representing the number of significant components. This condition enables CS algorithms to reconstruct the original signal even when the number of measurements is far lower than the signal’s dimensionality \citep{candes_introduction_2008}. 

\begin{figure}
\includegraphics[width=0.48\textwidth]{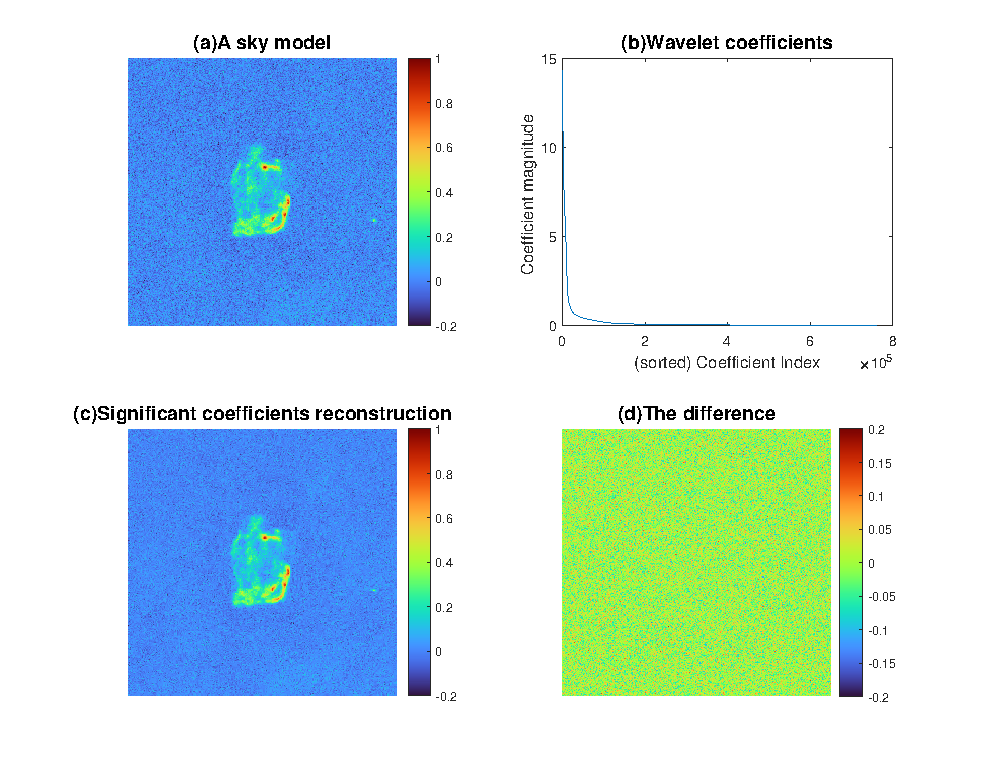}
\caption{A sky model and its reconstruction from significant IUWT coefficients.  $\mathbf{(a)}$: A noisy sky model, generated by adding Gaussian noise (with $\mu = 0$, $\sigma = 0.01$) to the normalized sky model ($\text{sky model}/ \max(\text{sky model})$). $\mathbf{(b)}$: The IUWT coefficients of the noisy sky model, sorted in descending order, where the magnitude of most coefficients is close to zero. The x-axis represents the coefficient index, while the y-axis indicates the magnitude. $\mathbf{(c)}$: The reconstruction from the significant coefficients of $\mathbf{(a)}$, with values below 0.012 of the maximum magnitude set to zero, resulted in 88.4\% of coefficients being zero. $\mathbf{(d)}$: The difference between the reconstruction and the noisy sky model, primarily representing the noise. Note that the level of the IUWT is 4 and the magnitude is an arbitrary unit.}

\label{fig:cwoef }
\end{figure}

In the context of radio interferometric deconvolution, wavelet and curvelet bases, are among the most commonly utilized methods for sparse representation (such as \cite{li_application_2011}, \cite{garsden_lofar_2015}, \cite{dabbech_moresane:_2015} , \cite{cai_uncertainty_2018} and \cite{muller_dog_hit_2022}). Fig. \ref{fig:cwoef } illustrates a sky model\footnote{The original model is available at: \url{https://casaguides.nrao.edu/index.php/Sim_Inputs}} and its reconstruction from significant isotropic undecimated wavelet transform (IUWT) coefficients. The example illustrates the effectiveness of the wavelet dictionary in representing the signal through significant coefficients, resulting in a sparse representation.

The concept of incoherence pertains to the lack of correlation between the measurement domain and the sparsity domain. It refers to how unrelated or orthogonal the sampling basis vectors are to the sparsity basis vectors. The coherence between the sampling matrix $\Phi$ and the representation basis $\Psi$ is \citep{candes_introduction_2008}:
\begin{equation}
\mu(\Phi,\Psi) = \sqrt{n}\cdot \max_{1\leq i,j\leq n}{|\langle\phi_i, \psi_j \rangle|}
\label{eq:mucohe}
\end{equation}
where $\phi_i$ is the $i$-th column of $\Phi$, and $\psi_j$ is the $j$-th column of $\Psi$.

Incoherence ensures that the measurement matrix captures diverse information about the signal, with each measurement contributing independently. This property facilitates the recovery of the original sparse signal from a limited number of measurements and mitigates the impact of noise on the reconstruction process, resulting in more accurate and stable recovery. For example, significant incoherence can be observed between the time and frequency domains \citep{foucart_mathematical_2013,schulz_compressive_2021}. Random matrices, such as Gaussian or Bernoulli matrices, and Fourier matrices have been shown to effectively satisfy the incoherence condition in many scenarios \citep{candes_introduction_2008}. When applying a Gaussian random as the sampling matrix, the number of required measurements $m$ satisfy the following condition \citep{foucart_mathematical_2013,wright_high_dimensional_2022}:
\begin{equation}
m \propto   k \cdot \log(n/k)
\label{eq:mknrealtion}
\end{equation}
This relationship underscores the interplay between the number of measurements $ m $, the signal dimension $ n $, and the $ k $ most significant coefficients. In terms of radio interferometric observations, $ m $ and $ n $ are typically known quantities, making the $ k $ the critical parameter influencing signal recovery.

In linear systems, to maintain the Euclidean norm of signals, the sensing matrix $ \Phi $ should  be orthogonal or quasi-orthogonal, which satisfies the following condition for any vector $ u $:
\begin{equation}
\|\Phi u\|_2^2 = \|u\|_2^2 \quad \forall u
\label{eq:orthphi}
\end{equation}
This condition indicates that the sensing matrix preserves the $ \ell_2 $-norm (Euclidean length) of any signal $ u $. However, in the presence of noise, preserving the Euclidean length of the signal becomes more challenging as noise can distort the measurement outcomes. To address this, the sensing matrix must satisfy the restricted isometry property. The matrix $ \Phi $ satisfies the RIP if, for any $ k $-sparse vector $ u $, the following holds approximately for $ \delta_k \in (0, 1) $ \citep{candes_introduction_2008}:
\begin{equation}
(1 - \delta_k) \| u \|_2^2 \leq \| \Phi u \|_2^2 \leq (1 + \delta_k) \| u \|_2^2
\label{eq:rip}
\end{equation}
RIP ensures that the sensing matrix preserves the distances between sparse signals, even in noisy environments, thereby enabling accurate and stable signal recovery. An equivalent description of the RIP is that all subsets of $\Phi$ columns are orthonormal or nearly orthonormal \citep{candes_introduction_2008}.

\section{The Solution by the Split Bregman Method}\label{sec:ssbm}

The equation (\ref{eq:consbv}) can be solved iteratively using the split Bregman method:
\begin{equation}
\begin{split}
rhs^k=\mu\mathcal{F}^{\dagger} M^T V_m+\lambda\nabla^T_x(d_x^k-b_x^k)+\lambda \nabla^T_y(d_y^k -b_y^k)\\
+\gamma \Psi^T(w^k -b_w^k)
\end{split}
\label{eq:s0}
\end{equation}

\begin{eqnarray}
\label{eq:s1}I^{k+1}=\mathcal{F}^{\dagger}[K^{-1}\mathcal{F}(rhs^k)] \\
\label{eq:s2}d_x^{k+1}=\max(s^k-\frac{1}{\lambda},0)\frac{\nabla_x I^k + b_x^k}{s^k}\\
\label{eq:s3}d_y^{k+1}=\max(s^k-\frac{1}{\lambda},0)\frac{\nabla_y I^k + b_y^k}{s^k}\\
\label{eq:s4}w^{k+1}=\mathscr{S}(\Psi I^{k+1} + b_w^k,\frac{1}{\gamma})\\
\label{eq:s5}b_x^{k+1}=b_x^k+(\nabla_x I^{k+1}-d_x^{k+1})\\
\label{eq:s6}b_y^{k+1}=b_y^k+(\nabla_y I^{k+1}-d_y^{k+1})\\
\label{eq:s7}b_w^{k+1}=b_w^k+(\Psi I^{k+1}-w^{k+1})\\
\label{eq:s8}V_m^{k+1}=V_m^k+V_m-M\mathcal{F}I^{k+1}
\end{eqnarray}
where $K=\mu M^TM-\lambda\Delta+\gamma \mathbf{I}$, $\mathbf{I}$ is the identity matrix and  $s^k = \sqrt{|\nabla_x I^k + b_x^k|^2+|\nabla_y I^k + b_y^k|^2}$. 
The $\mathscr{S}$ means the shrinkage(soft thresholding) operator, where $\mathscr{S}(\alpha,\beta) =\frac{\alpha}{|\alpha|}\max\{|\alpha|-\beta,0\}$. Note that $\nabla^T\nabla=-\Delta, \;\Psi^T\Psi=\mathbf{I}, \; \mathcal{F}^T=\mathcal{F}^{\dagger}$.

\section{The Reformulation of the Split Bregman Method}\label{sec:refsbm}

The  $\max$ operator in equations (\ref{eq:s2}) and (\ref{eq:s3}) can be seamlessly substituted with the ReLU operator, which in its reformulated state is equivalent to the shrinkage operator in equation (\ref{eq:s4}). Therefore, equations (\ref{eq:s2}) to (\ref{eq:s4}) become:

\begin{eqnarray}
\label{eq:ns2}d_x^{k+1}=\text{ReLU}(s^k-\frac{1}{\lambda^k},0)\frac{\nabla_x I^k + b_x^k}{s^k}\\
\label{eq:ns3}d_y^{k+1}=\text{ReLU}(s^k-\frac{1}{\lambda^k},0)\frac{\nabla_y I^k + b_y^k}{s^k}\\
\label{eq:ns4}w^{k+1}=\text{ReLU}(\Psi I^{k+1} + b_w^k-\frac{1}{\gamma^k})
\end{eqnarray}

This reformulation enables the split Bregman method to be reconstructed as a neural network. 


\section{The configurations of the Methods}\label{sec:m31mconfig}

For the MS-CLEAN method, we set a gain of $0.1$, with seven scales $[0, 2, 4, 8, 16, 32, 64]$, a $3 N_{\sigma}$ threshold, and a maximum of $10000$ iterations. The reweighted version implemented in MATLAB\footnote{Available at: \url{ https://code.google.com/p/csra/downloads}} is used for the IUWT-based CS method. We set the decomposition level to $5$, the threshold to $0.0001$, and the maximum number of iterations to $50$. For the MORESANE\footnote{Available at: \url{https://github.com/ratt-ru/PyMORESANE}} method, we initially took the parameters listed in their study. However, the reconstruction failed, prompting us to increase the threshold of the wavelet coefficients from $4N_{\sigma}$ to $9N_{\sigma}$ and reduce the loop gain from $0.2$ to $0.01$. Ultimately, the best reconstruction was achieved by setting the threshold of wavelet coefficients to $7N_{\sigma}$ with a loop gain of $0.01$. Other parameters remained as suggested in their paper. Specifically, the percentage of the maximum wavelet coefficient was set to $\tau =0.7$. The maximum number of iterations for the major and minor loops were fixed at 200 and 50, respectively. And the precision parameter is $0.0001$.

Since $V_m$ represents noisy measurements, our deconvolution method fundamentally involves estimating the true visibility $V_{\text{true}}$ from $V_m$. This estimation balances the trade-off between minimizing noise and maintaining fidelity to the original signal. The parameters configuration is introduced in section \ref{traind}.


\section{The Simulated observation of 3C 288 Radio Galaxy and Comparison Results}\label{sec:3c288}

\begin{figure}
\centering
\includegraphics[width=0.48\textwidth]{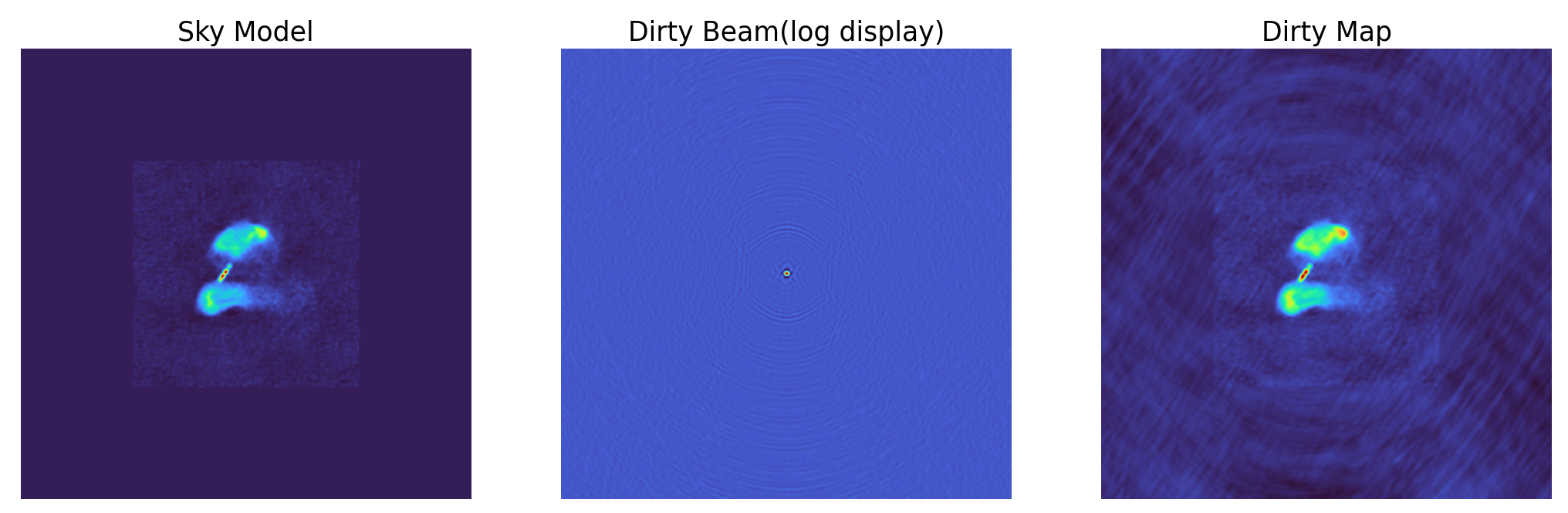}
\caption{\emph{Left}: the 3C\,288  radio galaxy model padding with zeros. \emph{Middle}: the 8\,h observation dirty beam displayed with logarithmic scaling. \emph{Right}: the observed dirty map. All images demonstrated the central regions.}
\label{fig:3c288-obs}
\end{figure}

The brightness of 3C\,288 radio galaxy model ranging from $-2.333 \times 10^{-4}$ to $0.013$ Jy/pixel. We normalized the image using the formula $ \text{image}_\text{normalization} =\text{image}/\max(\text{image})$, resulting in a normalized brightness range from $-0.0182$ to $1$ Jy/pixel. The configuration of the simulated observation is identical to that of the M31 experiment, except the observation time is extended to 8 hours and the noise standard deviation $N_{\sigma}$ is approximately $0.0122$ Jy/pixel. This results in a relatively high signal-to-noise ratio (SNR) compared to the M31 observation. The sky model image, the dirty beam, and the dirty map are shown in Fig. \ref{fig:3c288-obs}. 

\begin{figure*}
\centering
\includegraphics[width=0.7\paperwidth]{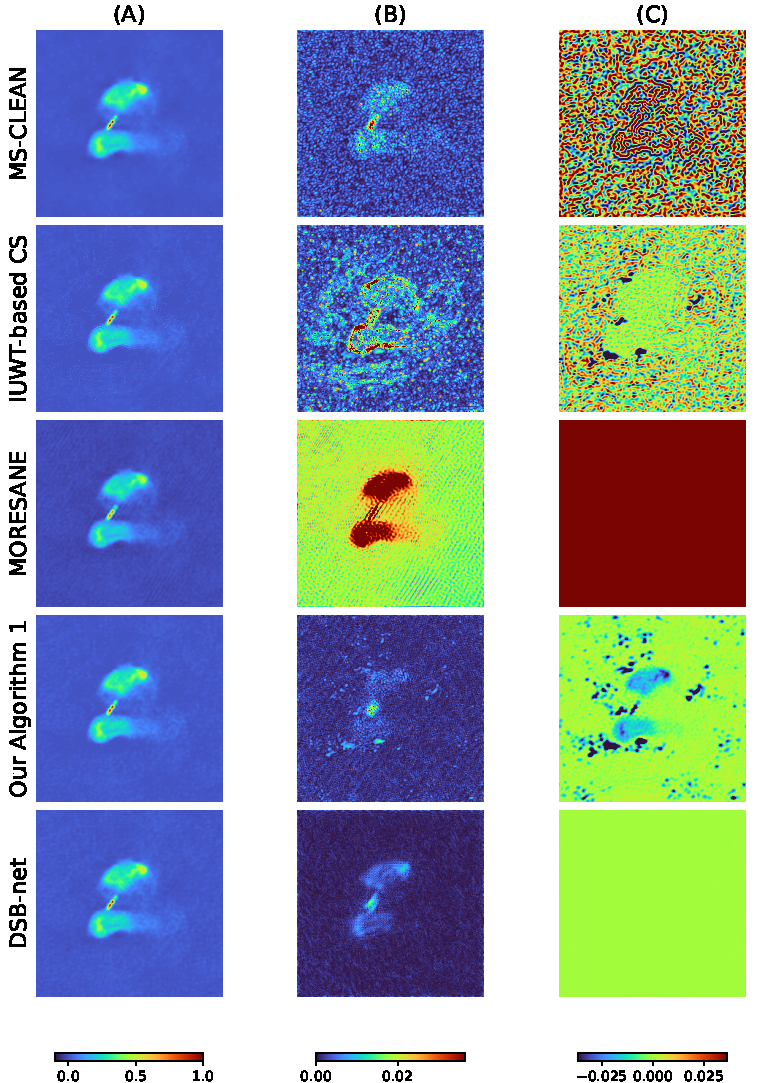}
\caption{The recovery images from the simulated  3C 288 radio galaxy observations with VLA. \emph{From top to bottom}: in rows are the results of the MS-CLEAN, the IUWT-based CS, the MORESANE method, the proposed Algorithm 1, and the DSB-net, respectively. \emph{From left to right}, $\mathbf{(a)}$ the model images, $\mathbf{(b)}$ model absolute error images (the absolute differences between the reference model images and the reconstructed model images), and $\mathbf{(c)}$ the residual images.}
\label{fig:3c288-compare}
\end{figure*}

\begin{figure*}
\centering
\includegraphics[width=0.8\paperwidth]{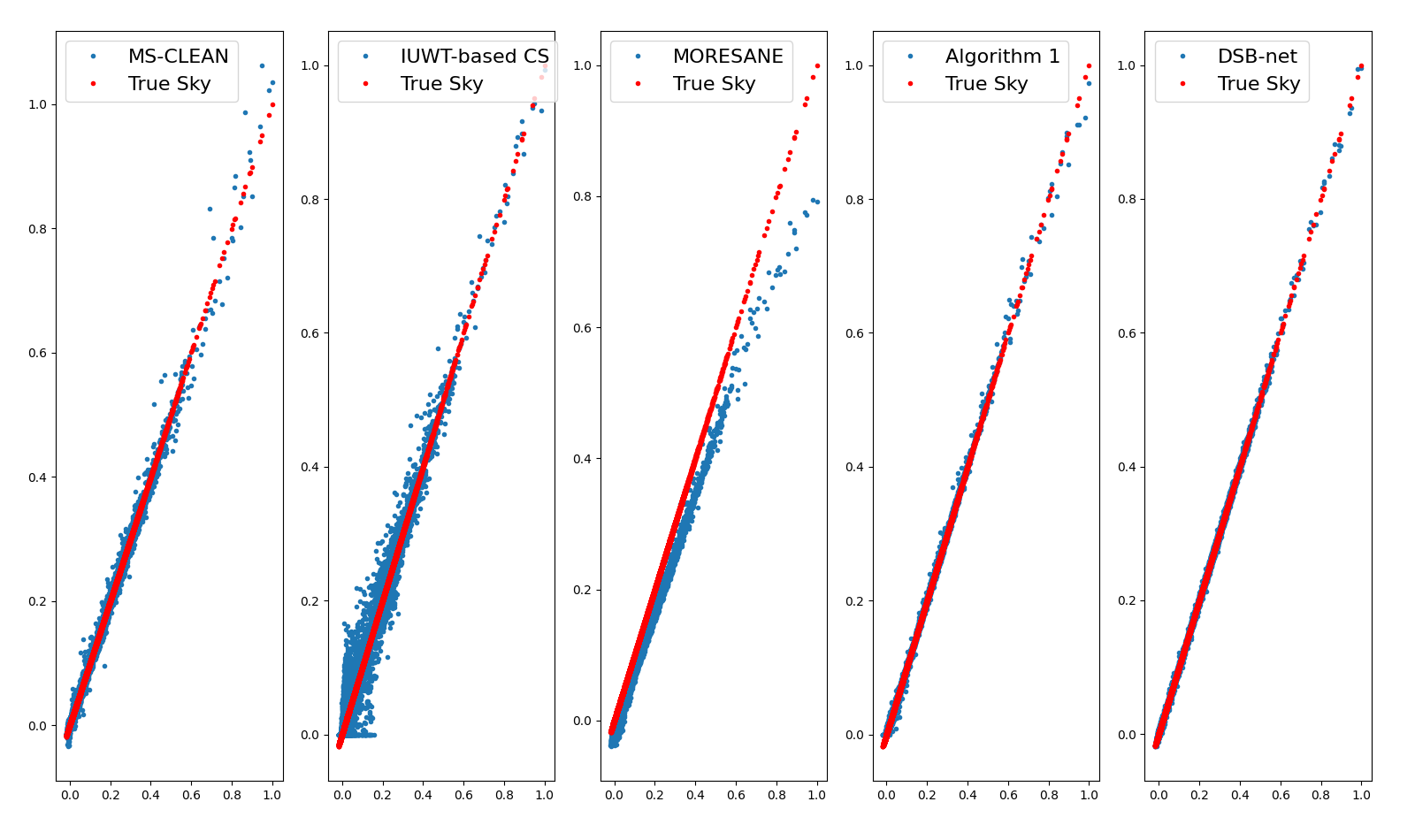}
\caption{The plots of the reconstructed models(y-axis) versus the true sky image model(x-axis). \emph{From left to right}: the recovery results through the MS-CLEAN, IUWT-based CS, MORESANE, the proposed Algorithm 1, and the DSB-net, respectively.}
\label{fig:3c288-compareTrue}
\end{figure*}

The parameter settings for the four deconvolution methods remain consistent with those used in the previous experiment, with the exception that the threshold for wavelet coefficients in MORESANE reverts to the authors' suggested value of $4N_{\sigma}$. The results are presented in Fig. \ref{fig:3c288-compare}, with the rows corresponding to MS-CLEAN, IUWT-based CS, MORESANE, the proposed algorithm 1 and DSB-net from top to bottom. The columns display the model images, model absolute error images, and residual images from left to right. The color bar ranges for the model absolute error images and the residual images are set to $0$ to $3N_{\sigma}$ and $-3N_{\sigma}$ to $3N_{\sigma}$, respectively. The reconstruction capability of the proposed DSB-net exceeds the other three deconvolution methods. Fig. \ref{fig:3c288-compareTrue} plots the recovered models against the true sky image, with the y-axis representing the four models and the x-axis representing the reference sky model. It is evident that the brightest sources of radiation in this sky model, which are the most challenging to recover accurately, are best reconstructed by DSB-net. Moreover, the numerical comparison of the recovery performance for the 3C\,288 radio galaxy is presented in Table \ref{tab:3c288ncom}. The results align with those observed for M31, demonstrating that our proposed DSB-net outperforms other methods. Specifically, the RMS error of DSB-net is approximately 1\% of that achieved by MS-CLEAN, while its DR is nearly 100 times higher than that of MS-CLEAN.

\begin{table*}
   \caption{The numerical comparison of the 3C\,288 radio galaxy reconstruction results. The robust weighting is applied to the UV coverage.} 
   \label{tab:3c288ncom}
    \centering
    \begin{tabular}{ c c c c c c}
\hline
  &MS-CLEAN&IUWT-based CS&MORESANE&   Algorithm 1 &DSB-net\\
\hline

RMS&0.0283&0.0168& 0.2049&  0.0107 &0.0003\\
DR&331&539&90& 786 &32606\\
\hline
    \end{tabular}
\end{table*}

\section{The influence of multi-resolution property}\label{sec:msinflu}

\begin{figure}
\centering
\includegraphics[width=0.48\textwidth]{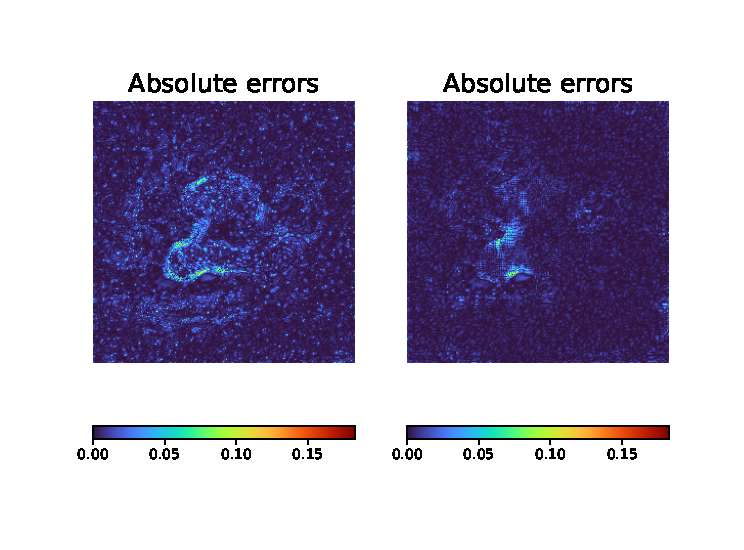}
\caption{\emph{Left}: the IUWT-based CS error image. \emph{Right}: the DWT-based CS error image. The error is the absolute difference between the reference model image and the reconstructed model image.}
\label{fig:iuwt-dwt}
\end{figure}

To demonstrate the impact of multi-resolution techniques on reconstructing the large-scale structure of the radio source, we applied the IUWT-based method to recover the 3C 288 sky model. However, in this instance, we replaced IUWT with DWT as the dictionary, thereby referring to it as the DWT-based CS method. The model absolute errors are illustrated in Fig. \ref{fig:iuwt-dwt}, with the color bar indicating an arbitrary range. The DWT-based CS method shows improved reconstruction results in the external structure of the lobes, which can be attributed to the interferometry array's short-spacing problem. The lower resolution facilitates easier prediction of the large-scale structure. Note that the decomposition level of both IUWT and DWT is set to $5$.

\section{The Simulated observation of point sources and Comparison Results}\label{sec:pskym}

\begin{figure}
\centering
\includegraphics[width=0.48\textwidth]{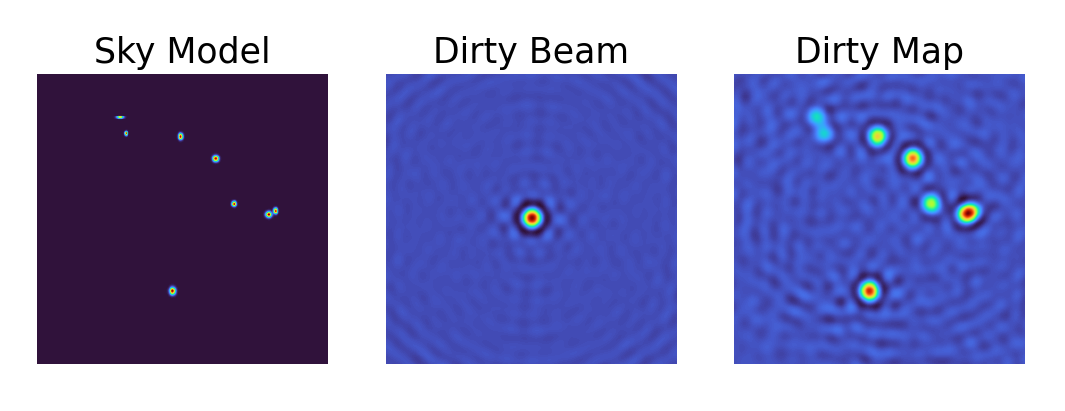}
\caption{\emph{Left}: the sky model padding with zeros. \emph{Middle}: the 8\,h observation dirty beam displayed with logarithmic scaling. \emph{Right}: the observed dirty map. All images demonstrated the central regions.}
\label{fig:psobs}
\end{figure}

An experiment featuring a sky model with dominant point sources and evaluates the performance of our method in comparison to MS-CLEAN. The observation configuration used the C configuration of the VLA and 8-hour observation time, while others remain consistent with that of the M31 simulated observation. The brightness of the sky model ranges from $1 \times 10^{-9}$ to 1.0 Jy/pixel. The sky model, dirty beam, and dirty map are presented in Fig. \ref{fig:psobs}, from left to right, respectively.

\begin{figure*}
\centering
\includegraphics[width=0.8\paperwidth]{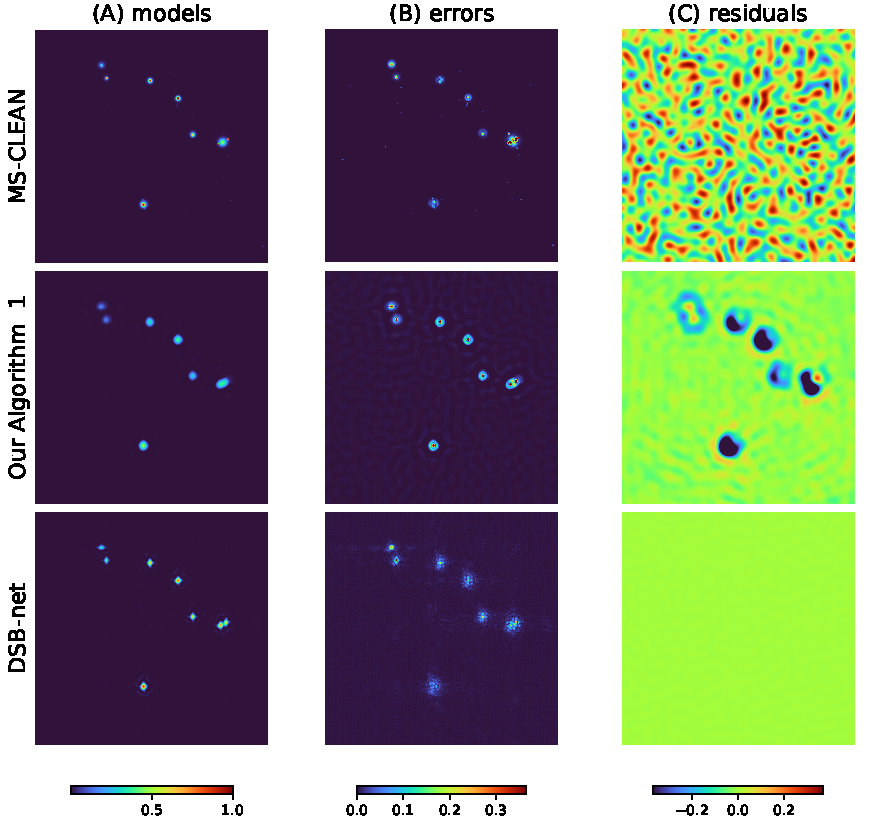}
\caption{The recovery images from the simulated point sources model. \emph{From top to bottom}: in rows are the results of the MS-CLEAN, the proposed Algorithm 1 and the DSB-net, respectively. \emph{From left to right}, $\mathbf{(a)}$ the model images, $\mathbf{(b)}$ model absolute error images (the absolute differences between the reference model images and the reconstructed model images), and $\mathbf{(c)}$ the residual images.}
\label{fig:psource-compare}
\end{figure*}

The reconstructed images from the observation of the simulated point sources are presented in Fig. \ref{fig:psource-compare}. From top to bottom in rows are the results of the MS-CLEAN, our method employing the DWT dictionary, and the proposed DSB-net, respectively. From left to right, (a) the model images, (b) model absolute error images (the absolute differences between the reference model images and the reconstructed model images), and (c) the residual images. The color bar range for the error images is set from 0 to  $3 N_{\sigma}$ while for the residual images, it is configured from  $-3 N_{\sigma}$ to $3 N_{\sigma}$.

The results suggest that, to some extent, our method demonstrates lower reconstruction accuracy compared to MS-CLEAN. While the RMS of the DSB-net method is slightly lower than that of MS-CLEAN (see Table \ref{tab:psncomp}), some artifacts, albeit below the $3 N_{\sigma}$ noise level, are present. The lower RMS value of DSB-net is primarily attributed to its superior effective angular resolution compared to MS-CLEAN, which aligns with findings from previous studies (e.g., \cite{garsden_lofar_2015,akiyama_superresolution_2017}). This improved resolution enables the distinction of two relatively close point sources.

\begin{table}
   \caption{The numerical comparison of the point sources model. \label{tab:psncomp}}
    \centering
    \begin{tabular}{ c c c c }
\hline
  &MS-CLEAN& Algorithm 1 &DSB-net\\
\hline

RMS& $7.03\times 10^{-5}$& $7.18\times 10^{-5}$ & $2.55\times 10^{-5}$\\
\hline
    \end{tabular}
\end{table}

\section{Computer configuration}

\begin{table*}
   \caption{Computer specifications for algorithm testing. \label{tab:comc}}
    \centering
   \begin{tabular}{ c c c}
\hline
  Part &Specification &Value\\
\hline
GPU& Nvidia GeForce RTX 4090& 24\,GB \\
CPU &AMD EPYC 7532  & 32-core $@$ 3.35GHZ3\\
Memory&DDR4  &512GB $@$ 3200MHz\\
\hline
 
\end{tabular}
\end{table*}

\end{appendices}


\bibliography{sample631}{}
\bibliographystyle{aasjournal}



\end{document}